\documentclass{article}
\usepackage{jinstpub}
\usepackage{graphicx} 
\usepackage{booktabs}
\usepackage{subcaption}
\usepackage[backend=biber,sorting=none]{biblatex}
\usepackage{lipsum}

\newcommand*{\affmark}[1][*]{\textsuperscript{#1}}

\title{Silicon pinhole strip defects and their impact on ATLAS Inner Tracker HV current measurement}
\author{Anthony~Affolder{\normalfont\it\affmark[a]},}
\author{Kirsten~Affolder{\normalfont\it\affmark[a]},}
\author{Emily Duden{\normalfont\it\affmark[b,]}
{\normalfont\affmark[1]},\note{Corresponding author}}
\author{Vitaliy~Fadeyev{\normalfont\it\affmark[a]},}
\author{Cole~Helling{\normalfont\it\affmark[c]},}
\author{David~Lynn{\normalfont\it\affmark[d]},}
\author{Forest~Martinez-Mckinney{\normalfont\it\affmark[a]},}
\author{Peter~Phillips{\normalfont\it\affmark[e]},}
\author{Luise~Poley{\normalfont\it\affmark[f,g]},}
\author{Tate~Sakaguchi{\normalfont\it\affmark[a]},}
\author{Abdullah~Sayed{\normalfont\it\affmark[b]},}
\author{Stefania~Stucci{\normalfont\it\affmark[d]},}
\author{Alex~Wang{\normalfont\it\affmark[a]},}
\author{Marcus~Wong{\normalfont\it\affmark[a]}}

\affiliation[a]{Santa Cruz Institute for Particle Physics, University of California at Santa Cruz, 1156 High Street, Santa Cruz, CA, USA}
\affiliation[b]{Brandeis University, Waltham, MA, USA}
\affiliation[c]{University of British Columbia, Agricultural Road, Vancouver, Canada}
\affiliation[d]{Brookhaven National Laboratory, Upton, NY, USA}
\affiliation[e]{Rutherford Appleton Laboratory, Didcot, UK}
\affiliation[f]{Department of Physics, Simon Fraser University, University Dr W, Burnaby, Canada}
\affiliation[g]{TRIUMF, Wesbrook Mall, Vancouver, Canada}

\date{April 2025}

\emailAdd{emily.rose.duden@cern.ch}

\begin{document}

\begingroup
\renewcommand\thefootnote{} 
\footnotetext[0]{%
\footnotesize
Copyright 2025 CERN for the benefit of the ATLAS ITk Collaboration. 
Reproduction of this article or parts of it is allowed as specified in the CC-BY-4.0 license.
}
\endgroup

\maketitle

\begin{abstract}
    \noindent\textsc{}In preparation for the High-Luminsoity LHC (HL-LHC)\cite{Aberle:2749422}, the ATLAS detector will undergo major detector upgrades, including the replacement of the current Inner Detector with the new all-silicon Inner Tracker (ITk)\cite{CERN-LHCC-2017-005}. The ITk consists of a pixel detector close to the beamline surrounded by a large-area strip detector. During detector production, the electrical properties of silicon sensors and readout electronics must be characterized through a series of quality control (QC) and quality assurance tests. These tests ensure any defect is captured at the earliest possible stage. One such defect, called a pinhole, occurs when the strip implant and the metal readout electrode are shorted through the intermediary dielectric layer. Notably, the introduction of pinholes during module assembly and pinhole effects on completed modules, especially on leakage current measurement circuitry, have never been studied. In this paper, we investigate the effect of such connections on the sensor leakage current measurements of completed modules and introduce new ways to locate pinholed strips. With minor modifications to testing procedures, such defects are shown not to impede module testing or performance.
\end{abstract}

\section{Introduction}

\subsection{The HL-LHC}

The HL-LHC will achieve a peak instantaneous luminosity of $7.5\times10^{34}$ cm$^{-2}$s$^{-1}$, corresponding to approximately 200 proton-proton collisions per bunch crossing, more than three times the current pileup. Subsequent radiation damage, limitations from detector occupancy, and bandwidth saturation will render the current ATLAS Inner Detector inoperable. As a result, the Inner Detector will be replaced with the ATLAS Inner Tracker, an all-silicon detector that consists of a pixel detector close to the beamline surrounded by a strip detector.

\subsection{ITk strip modules}

\begin{figure}

\begin{subfigure}{.4\textwidth}
  \centering
  \includegraphics[width=1.1\textwidth]{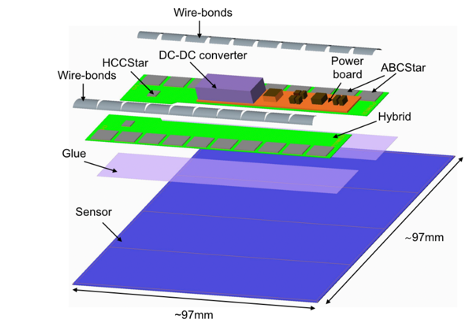} 
  \caption{\centering}
  \label{fig:SSModule}
\end{subfigure}%
\hspace{0.5cm} 
\begin{subfigure}{.4\textwidth}
  \centering
  \includegraphics[width=1.1\textwidth]{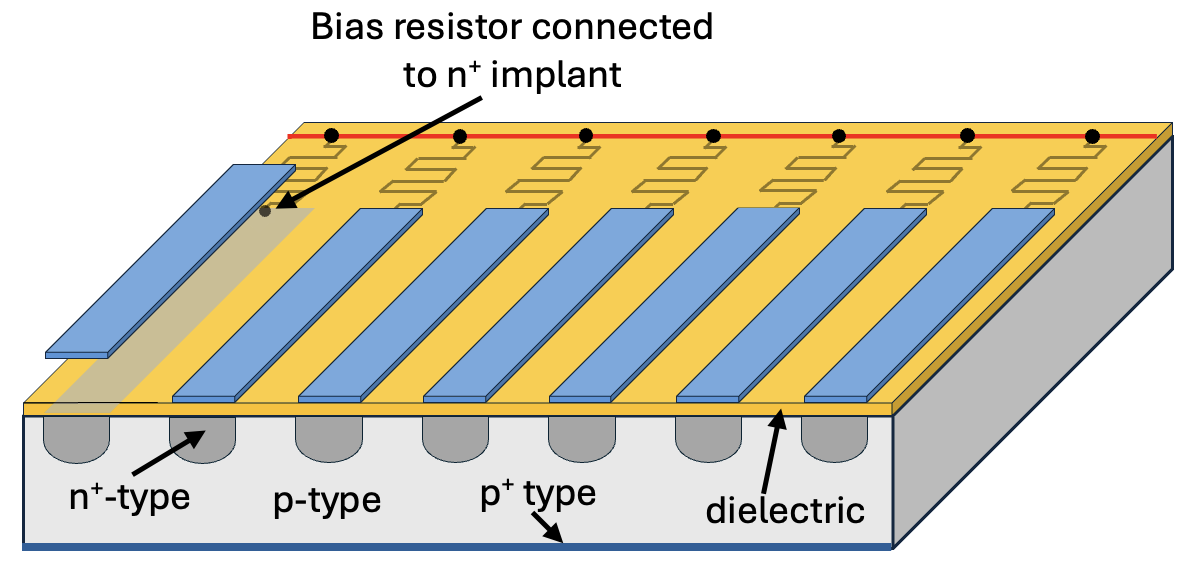}
  \caption{\centering}
  \label{fig:Silicon}
\end{subfigure}

\vspace{6pt}
\caption{(a) Exploded view of an ITk short-strip barrel module \cite{CERN-LHCC-2017-005}. (b) Cross section of an ITk strip sensor. Aluminum strips (blue) are AC-coupled via dielectric (yellow) to n$^+$ implant strips in a p-type bulk. Leakage current travels through the implant strips into bias resistors connected to the bias ring (red).}
\label{fig:modules}
\end{figure}

ITk strip modules consist of a silicon sensor with front-end electronics and powering and control circuitry glued on top (Figure \ref{fig:SSModule}). These sensors are AC-coupled strip sensors with n$^+$ implants coupled to a p-type bulk through a dielectric layer \cite{sensordesigns} (see Figure \ref{fig:Silicon}). A high voltage bias depletes the sensor and the resulting leakage current is collected by the bias ring.

\unskip

Several different module geometries exist in the strip detector. In the barrel region, short-strip modules with four rows of 1280 silicon strips make up two cylindrical layers closest to the beamline while long-strip modules with two such rows make up the surrounding lower-occupancy regions. These barrel modules use square $9.8\times9.8$ cm$^2$ sensors with 75.5 $\mu$m strip pitch, while endcap sensors have approximately trapezoidal geometries with curved edges and a 70 to 80 $\mu$m strip pitch necessary to create a hermetic circular endcap disk \cite{LACASTA2019137}. All strip modules employ the same Application Specific Integrated Circuits (ASICs) for signal readout in the front-end, powering, and control. 

\begin{figure}
\centering
\hspace{-12pt}\includegraphics[width=10 cm]{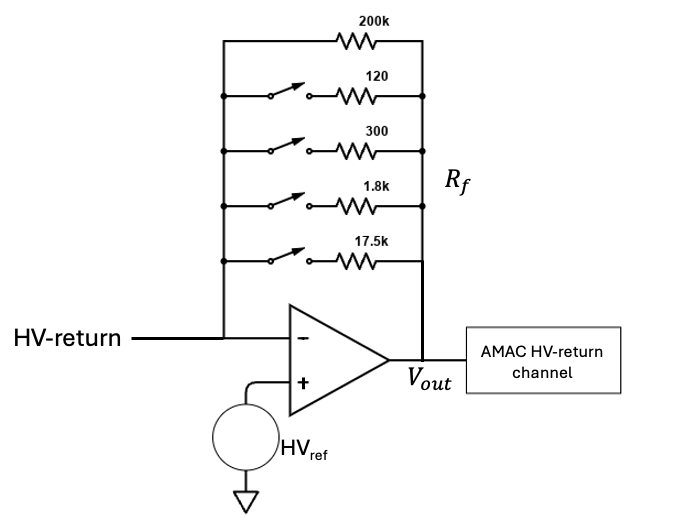}

\caption{AMAC circuit for measuring leakage current traveling through the HV-return line. }
\label{fig:AMAC}
\end{figure}

Power for front-end electronics and monitoring and control functionality is provided by the power board. Power boards receive 11V input voltage and output 1.5V to front-end chips via an aluminum shielded buck converter called the DCDC \cite{Faccio:2020DV}. Control of module electronics and monitoring of the module environment via measurements of voltage, current, and temperature are performed by the Autonomous Monitor and Control (AMAC) chip \cite{AMAC_Gosart}, also hosted by the power board. The bias ring is connected to the AMAC for leakage current measurements. The AMAC HV-return measuring circuit is shown in Figure \ref{fig:AMAC}. An operational amplifier (op amp) receives the leakage current from the HV-return line and outputs a voltage. Note that the feedback includes a permanent $200$ k$\Omega$ resistor with other optional feedback resistors that can be put in parallel, lowering the feedback resistance. Output voltage is then digitized into ADC counts by the AMAC, which can be converted into leakage current using feedback resistance, a mV/counts calibration factor (usually close to 1), and 0V offset (ADC counts when no sensor bias voltage is applied). At maximum, the AMAC outputs 1023 ADC counts. A typical 0V offset is $\sim$100 ADC counts, corresponding to 100 mV voltage at the non-inverting op amp input.

The front-end consists of several ATLAS Binary Chips (ABCs) \cite{ABC_Cormier} which amplify, shape, and discriminate signal before sending it to a Hybrid Controller Chip \cite{HCC_Dandoy} for packaging and shipping out of the module. Each ABC has 256 channels which are directly bonded to individual strips on the sensor.

Completed modules are glued onto lightweight carbon fiber support ``cores'' to form units for insertion into the ITk detector. These units, called staves in the barrel and petals in the endcaps, provide mechanical support, cooling, and electrical services to modules. The relatively simple geometry of the barrel permits rectangular staves with 14 modules per side. Endcap petals are wedge-shaped and consist of six modules per side.

During module and stave production, rigorous QC tests are performed \cite{AbeQC,ExtremeQC}. These include IV scans, where sensor leakage current is measured with the AMAC as a function of bias voltage. IV scans are essential to revealing sharp increases in current before the -500V reverse bias voltage specification, called early electrical breakdown. Room temperature per-strip leakage current measurements are $<0.1$ nA during QC (before irradiation in HL-LHC). Per-strip front-end measurements are also performed during QC, where charges of 0.5 fC, 1.0 fC, and 1.5 fC are internally injected in each ABC channel and the discriminator threshold is scanned, resulting in characteristic ``S-curves'' of signal efficiency. The point of crossing 50$\%$ efficiency is termed $V_T(50)$ and the transition is a measure of overall noise. Channel gain in mV/fC is derived from the relationship between $V_T(50)$ and injected charge. Overall noise and channel gain are used to calculate strip input-referred noise.

\subsection{Electrical testing and ``pinhole'' strip defects}

Connections between the silicon strip implant and the readout electrode are expected to arise, albeit rarely, during sensor and module production. These connections, called pinholes \cite{Hartmann2024-zo}, provide a path for leakage current to enter front-end electronics, potentially affecting front-end chip operations. Pinholes presented a recurring challenge in prior silicon detector experiments \cite{AXER2004321, FADEYEV2020163991}. Fortunately, pinholes in ITk strip modules have been previously studied \cite{Affolder_2021} and shown not to affect front-end measurements up to 250 nA current per strip, beyond the expected end-of-life leakage current of the detector. However, the effect of pinhole-like connections on the HV-return circuit has never been investigated. Since pinholes not only allow leakage current to enter the front-end, but also connect front-end electronics to the HV-return line, AMAC leakage current measurements can be affected by channels with pinholes. This compromises the ability of the AMAC to accurately measure the leakage currents and identify early breakdown in the sensor. Thus, it is of paramount importance to understand and reduce the effect of pinholes on AMAC measurements, and identify the location and causes of pinholes whenever possible.

\section{Modeling pinhole-like connections with the HV return}
\label{sec:modeling}

\subsection{A simplified ITk module circuit}
\begin{figure}
\centering
\hspace{-12pt}\includegraphics[width=10 cm]{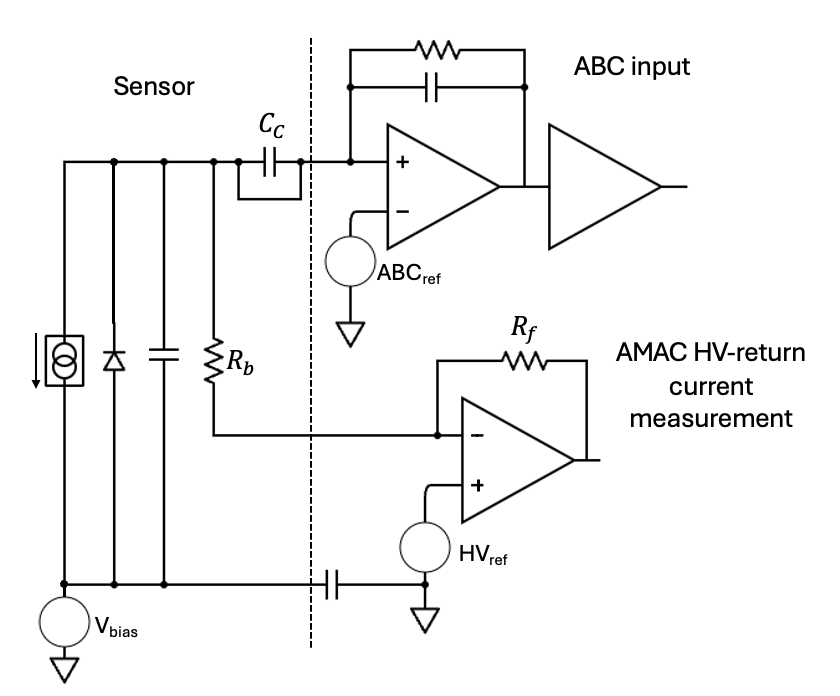}

\caption{Model of a strip, ABC channel input, and AMAC HV-return channel. A pinhole is represented as a short across the dielectric, represented by $C_C$. $R_b$ is the bias resistor.}
\label{fig:AMACABC}
\end{figure}

A pinhole can be modeled as a short bypassing the dielectric layer, represented in Figure \ref{fig:AMACABC} by a shorted capacitor $C_C$. If the pinhole connection is indeed a short, the only impedance between ABC input and AMAC input comes from the bias resistors that connect the implant strip to the bias ring. With the ABC input directly connected to the AMAC, direct current may now flow between the two chips when there exists a potential difference. Indeed, the ABC's input voltage with nominal settings is $ABC_{ref}=$ 250 mV while the AMAC reference voltage is $HV_{ref}=115$ mV. 

Note that this model describes only a single strip and ABC input channel. In reality, multiple implant strips (and multiple metal strips when more than one pinhole is present) are connected to the HV-return line. However, we can understand the effect of many strips in parallel on AMAC HV-return measurements by connecting the ABC input voltage $ABC_{ref}$ with the HV return line via an impedance $R_h$, which in the case of pinholes is equal to:

\begin{equation}
    R_h = \frac{R_S+R_b}{\# pinholes}
\end{equation}

where $R_S$ is the impedance of a pinhole (if it is indeed not a short). The bias resistors have values of $R_b=1.5\pm 0.5$ M$\Omega$. The model of the AMAC HV current measuring circuit that includes a connection to an outside voltage (in this case, from ABC via a pinhole) is equivalent to the AMAC current measuring circuit shown in Figure \ref{fig:AMAC} but with the addition of an outside voltage source $V_S$ connected to the HV-return line via $R_h$. Using the ideal amplifier approximation, one can easily calculate the output voltage of the circuit in the absence of leakage current, which is subsequently converted into ADC counts inside the AMAC:

\begin{equation}
    V_{out}= \frac{R_f}{R_h}(HV_{ref}- V_S)+HV_{ref}
    \label{eq:Vout}
\end{equation}

Equation \ref{eq:Vout} is equivalent to the 0V offset (since it holds when no bias voltage is applied) and that the measured offset is affected by outside voltages connected to the HV-return. Note that this equation holds for the general case of any external voltage connection the the HV-return line. For example, should the HV line be shorted to ground, we see that $R_h = 0$ and $V_{out}\rightarrow \infty$. During sensor testing before module assembly, the HV biasing circuit is completed by connecting the bias ring to ground on the testing frame and occasionally this bond is mistakenly left in place when the module is completed. Indeed, we observe AMAC ADC values of 1023 when this occurs---the AMAC op amp has saturated high.

In the case of ABC connection to the AMAC via pinholes, the op amp output voltage will be pulled down, since $V_S = ABC_{ref}>HV_{ref}$. Should multiple pinholes exist, the pinhole connection have sufficiently low impedance, or ABC input voltage be suitably high, this can result in the AMAC saturating low.

\subsection{Consequences for electrical tests}

With this simplified model of pinholes, one can consider two things: 1) pinhole effects on AMAC measurements and 2) pinhole effects on ABC measurements. The latter has been largely explored by Affolder et al. \cite{Affolder_2021}, where it was shown that pinholes do not affect ABC measurements until the per strip leakage current exceeds 250 nA. At this current level, channels with pinholes exhibit noticeable losses in the gain compared to non-pinhole channels, and one can identify individual channels with pinholes by observing this reduced gain. QC tests on front-end electronics are thus unaffected by pinholes due to very small values of leakage current. This leaves AMAC measurements as the most straightforward means of identifying and studying pinholes.

As shown in Equation \ref{eq:Vout}, the first hint of a pinhole is evident in a decrease of the HV-return ADC counts. With the AMAC reference voltage of approximately 100 mV and the ADC conversion factor of $\sim$1 count/mV, we expect $\sim$100 ADC counts when no bias voltage is applied. Due to statistical fluctuations, this value varies by several counts between measurements. Pinholes may pull the counts down below 100. When the pinhole connection is sufficiently strong, i.e. pinhole resistance is low or several pinholes are present, the AMAC saturates low and the counts will not fluctuate, but instead remain constant at a low value. 

This has significant consequences for IV tests of constructed modules, where AMAC ADC counts are directly converted into current measurements. When the sensor is biased, current flows between the HV return and op amp, and an additional term for this current must be added to Equation \ref{eq:Vout} to calculate AMAC output voltage. The bias voltage at which this current is sufficient to overcome AMAC saturation is determined by pinhole strength. This means IV scans taken on modules with pinholes can exhibit an initially unchanging current that suddenly increases, or a constant, zero current. Fortunately, one can easily obtain unsaturated current measurements by turning off ABC power via the DCDC buck converter, which sets ABC input voltage to ground. In this case, the ADC counts offset is increased by the connection to a voltage lower than $HV_{ref}$, but changes in current and breakdown behavior as bias voltage increases can still be observed. 

From Equation \ref{eq:Vout}, we can also infer that decreasing the AMAC feedback resistance $R_f$ will also lessen the effect of pinholes. If $R_f$ is sufficiently lowered, this will move the AMAC out of the saturated regime. 

\section{Observations of pinhole-like connections on ITk modules}

\subsection{Brookhaven National Lab (BNL)}

\subsubsection{On barrel modules}

Several modules have been observed with pinholes at BNL. Notably, four modules were extensively studied in the summer of 2023. These modules were first flagged during module QC, when the IV scans showed no change in current up to a 550V reverse bias voltage. Their ADC counts clearly saturated low; the counts at default feedback resistance ($200$ k$\Omega$) and 0V bias are given in Table \ref{tab:initialcounts}. Decreasing the feedback resistance on these modules eliminated the saturation, bringing the ADC counts at 0V bias close to the expected offset (see Table \ref{tab:countsbyRf}). In all cases, the AMAC stopped saturating between a feedback resistance of $200$ k$\Omega$ and $16$ k$\Omega$. This reduced impact of pinholes on ADC counts with decreasing feedback resistance is predicted by Equation \ref{eq:Vout}, supporting the pinhole model described in Section \ref{sec:modeling}.

\begin{table}[h]
    \centering
    \begin{tabular}{lccc}
        \toprule
        MLS-224 & MLS-236 & MLS-240 & MLS-244 \\  
        \midrule
        51 & 47 & 51 & 45 \\  
        \bottomrule
    \end{tabular}
    \caption{Initial AMAC HV-return ADC counts with $R_f=200$ k$\Omega$ and 0V bias for four modules built and tested at BNL. All counts do not fluctuate between consecutive measurements and are significantly lower than the $\sim$100 count offset expected, indicating pinholes are causing the AMAC on these modules to saturate low.}
    \label{tab:initialcounts}
\end{table}

\begin{table}[h]
    \centering
    \begin{tabular}{l|cc}
        \toprule
        {$R_f$ ($\Omega$)} & MLS-240 & MLS-244 \\  
        & ADC counts & ADC counts \\
        \midrule
        200k  & 51  & 45  \\  
        16k  & 110 & 105  \\  
        1.8k  & 140 & 135  \\  
        0.3k  & 144 & 135  \\  
        \bottomrule
    \end{tabular}
    \caption{AMAC HV-return ADC counts for two BNL modules with 0V bias at different AMAC feedback resistances. Between $R_f$ values of $200$ k$\Omega$ and $16$ k$\Omega$, both modules stop saturating low and show counts close to those expected for 0V offset. Consecutive measurements under the same conditions vary by a few counts, though this fluctuation is much smaller than the effect of the $R_f$ change.}
    \label{tab:countsbyRf}
\end{table}

To verify the source of the AMAC saturation, several checks were done. First, ADC counts were measured with the DCDC conversion circuit (which powers the front-end) on and off. The counts increased significantly when the DCDC was turned off, confirming that a voltage originating from the DCDC was connected to the HV return line. To determine the location of this connection, the bond connecting the bias ring to the rest of the HV return line was removed, effectively removing the sensor load from the HV-return. The subsequently measured counts agreed with expectations at 0V bias, confirming that a voltage was introduced to the HV return via the sensor. AMAC ADC counts at 0V bias on one representative module in various scenarios are given in Table \ref{tab:countsdiffconditions}. Since the only device the DCDC powers in proximity to the sensor are the ABCs, the problem was identified as a connection between strip and ABC input voltage via a pinhole. 

\begin{table}[h]
    \centering
    \begin{tabular}{l|cc}
        \toprule
        DCDC on/off & Other conditions & ADC counts \\ 
        \midrule
        On  & nominal & 45  \\  
        Off  & nominal & 240  \\  
        On  & bias ring/HV-return bond cut & 137  \\  
        \bottomrule
    \end{tabular}
    \caption{AMAC HV-return ADC counts for BNL-PPB-MLS-244. Counts with DCDC on and off agree with expectations for a pinholed module. Removing the sensor load gives typical non-pinhole module counts even with the DCDC on. Again, these values represent only single measurements and can vary by several counts between measurements.}
    \label{tab:countsdiffconditions}
\end{table}

\begin{figure}
\centering
\includegraphics[width=0.5\textwidth]{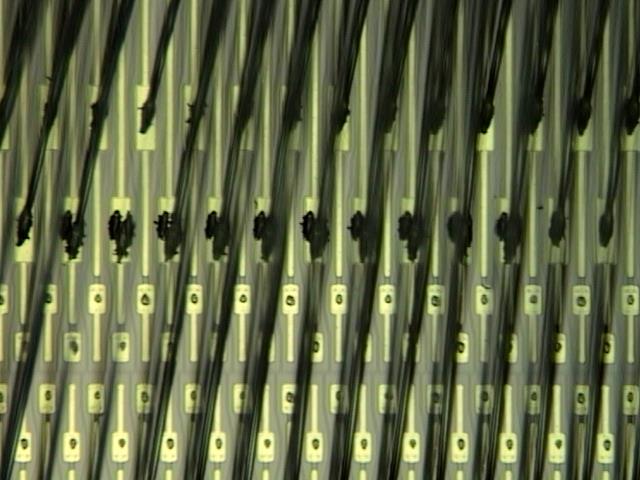}

\caption{Microscope image of several channels on module BNL-PPB-MLS-244 where bonds were removed before re-bonding. The channels were initially bonded off-center, making contact with the silicon passivation layer and potentially causing a pinhole connection.
}
\label{fig:BNLwirebond}
\end{figure}

In this case, the cause of pinholes was likely the damage of the dielectric layer near the aluminum strips during bonding. In all four of these modules, microscope images indicated mistakes in bonding. An example is shown in Figure \ref{fig:BNLwirebond}. Given the extensive bonding issues on these modules, it is unsurprising that pinholes were observed

\subsubsection{On an endcap module}

\begin{figure}

\begin{subfigure}{.5\textwidth}
  \centering
  \includegraphics[width=\textwidth]{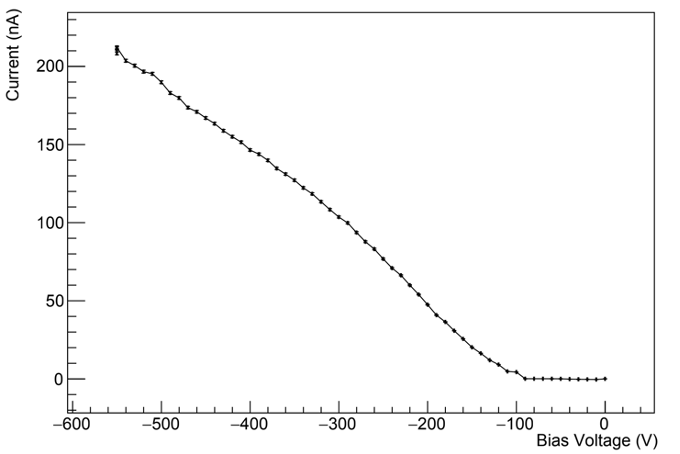}
  \caption{\centering}
  \label{fig:EndcapDCDCOn}
\end{subfigure}%
\begin{subfigure}{.5\textwidth}
  \centering
  \includegraphics[width=\textwidth]{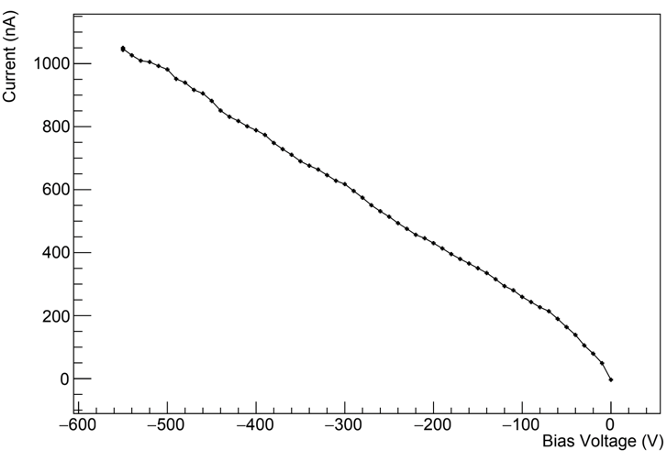}
  \caption{\centering}
  \label{fig:EndcapDCDCOff}
\end{subfigure}
\vspace{6pt}
\caption{IV curves taken on an endcap module (TRIUMF$\_$R5$\_$0007) built in Vancouver and tested at BNL with (a) ABCs powered and (b) ABCs unpowered. Though with powered ABCs the module appears to exhibit a sudden increase in current at 100V characteristic of early breakdown, the current is actually constant at 0 nA prior to this increase, indicating the AMAC was simply saturating low at lower bias voltages. This behavior disappears when ABCs are switched off.}
\label{fig:endcapmod}
\end{figure}

While BNL is a barrel module and stave production site, in Fall 2023 an endcap module built in Vancouver called TRIUMF$\_$R5$\_$0007 was sent to BNL for studies related to an early breakdown supposedly observed in the module. An IV scan was performed on this module with the ABCs powered (Figure \ref{fig:EndcapDCDCOn}) in which a sharp increase in current is observed at 100V. This module was initially believed to have an early breakdown at 100V bias voltage because of the increase. However, direct AMAC ADC count measurements revealed that the AMAC simply saturated low at bias voltages $<$100V, and the sudden increase was the result of the AMAC unsaturating as leakage current increased. A subsequent measurement with unpowered ABCs (Figure \ref{fig:EndcapDCDCOff}) further showed that the sensor did not exhibit early breakdown behavior and instead exhibits the common symptoms of pinholes. Additionally, $\sim$100 AMAC counts were measured at a lowered AMAC feedback resistance of $R_f = 16$ k$\Omega$, the expected value for a healthy module. Since this behavior---low saturation of the AMAC at low bias voltages only when the ABCs are powered and feedback resistance is high---has only been observed on modules with pinholes, the presence of pinholes in this module is highly likely.

\unskip

\subsubsection{On staves}

\begin{figure}
\label{fig:Jsidecounts}
\begin{subfigure}{.5\textwidth}
  \centering
  \includegraphics[width=\textwidth]{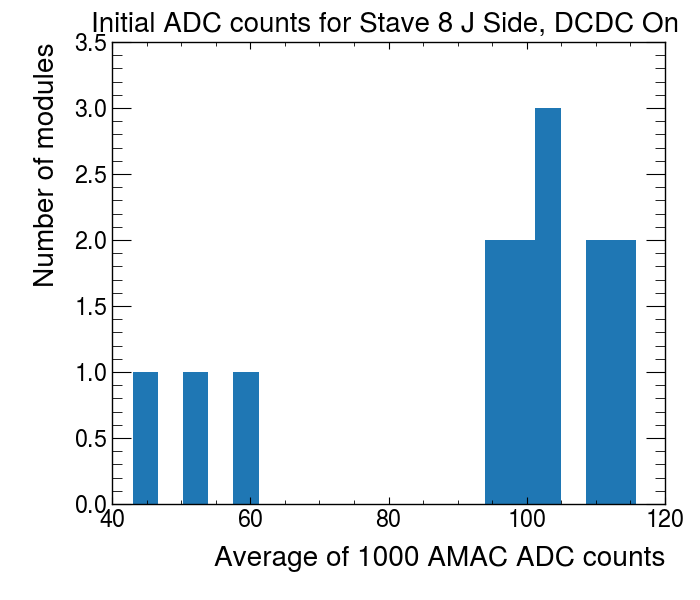}
  \caption{\centering}
  \label{fig:JsidecountsDCDCOn}
\end{subfigure}%
\begin{subfigure}{.5\textwidth}
  \centering
  \includegraphics[width=\textwidth]{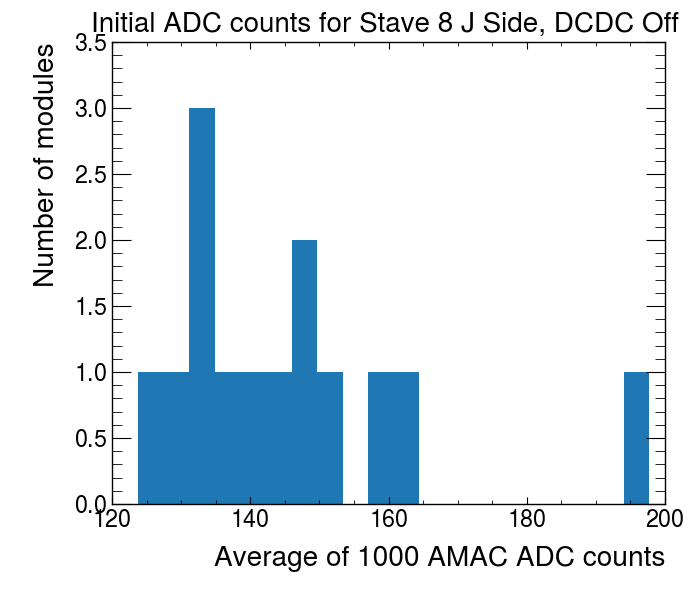}
  \caption{\centering}
  \label{fig:JsidecountsDCDCOff}
\end{subfigure}
\vspace{6pt}
\caption{Average of 1000 AMAC ADC counts for the 14 modules on the J side of Stave 8 taken immediately after loading, with (a) ABCs powered and (b) ABCs unpowered. Three modules show uncharacteristically low counts with ABCs powered, indicating those modules have pinholes.}
\label{fig:endcapmod}
\end{figure}

At BNL, the same electrical QC is performed on completed staves after module loading. Pinholes have also been observed on several staves at BNL. On one stave, called Stave 8, the AMAC saturation was tracked over time. None of the modules exhibited abnormal ADC counts at 0V bias before loading. The distribution of AMAC counts at 0V and ABCs powered on one side of Stave 8 after loading is shown in Figure \ref{fig:JsidecountsDCDCOn}. Three modules exhibit counts drastically reduced from the expected value, suggesting pinholes were introduced during the loading process. When ABCs are unpowered, AMAC counts from these modules with pinholes are inflated, as shown in Figure \ref{fig:JsidecountsDCDCOff}. This introduction of pinholes is most likely due to the tooling that is used to place modules onto the stave, which was subsequently improved.

A major issue facing module and stave production is the formation of cracks on sensors during the thermal cycling of the stave \cite{IAKOVIDIS2024169241, Tishelman-Charny:2024Hh}. While several cracking remediation strategies have been investigated and some were proven effective in solving the issue, the understanding of pinholes and their relation to cracking can be a useful diagnostic tool.

In Spring 2024, a stave undergoing thermal cycling to -45$^\circ $C exhibited a spike in current during a cold IV test. Subsequent electrical tests at room temperature showed early electrical breakdown and several channels with uncharacteristic noise, telltale signs of cracking. Thus, this current spike was determined to correspond to the moment of cracking. AMAC counts before and after the moment of cracking were studied to understand whether a pinhole was introduced. While the cracking occurred during an IV scan with powered ABCs, the time between voltage measurements was sufficient to measure average counts before and after the current spike at a constant bias voltage of 280V. After the current spike, AMAC counts increased by approximately 20 counts, characteristic of a pinhole effect when the ABCs are unpowered. After the IV test and an electrical test completed, the bias voltage was ramped to 0V with the ABCs powered. At 0V, the AMAC ADC counts remained unchanging at approximately 50. This low saturation with powered ABCs and elevated counts with unpowered ABCs indicate a pinhole was formed during the sensor cracking. While this observation requires constant monitoring of AMAC-measured current at the time of crack formation and has not been studied in subsequent crackings, it is useful nonetheless to understand that a sensor deformation from cracking can introduce a pinhole and the associated effects on HV-return measurements.

\subsection{Santa Cruz Institute for Particle Physics}

\begin{figure}[htbp]
  \centering
  \subfloat[Pinhole Module]{%
    \includegraphics[width=0.525\textwidth]{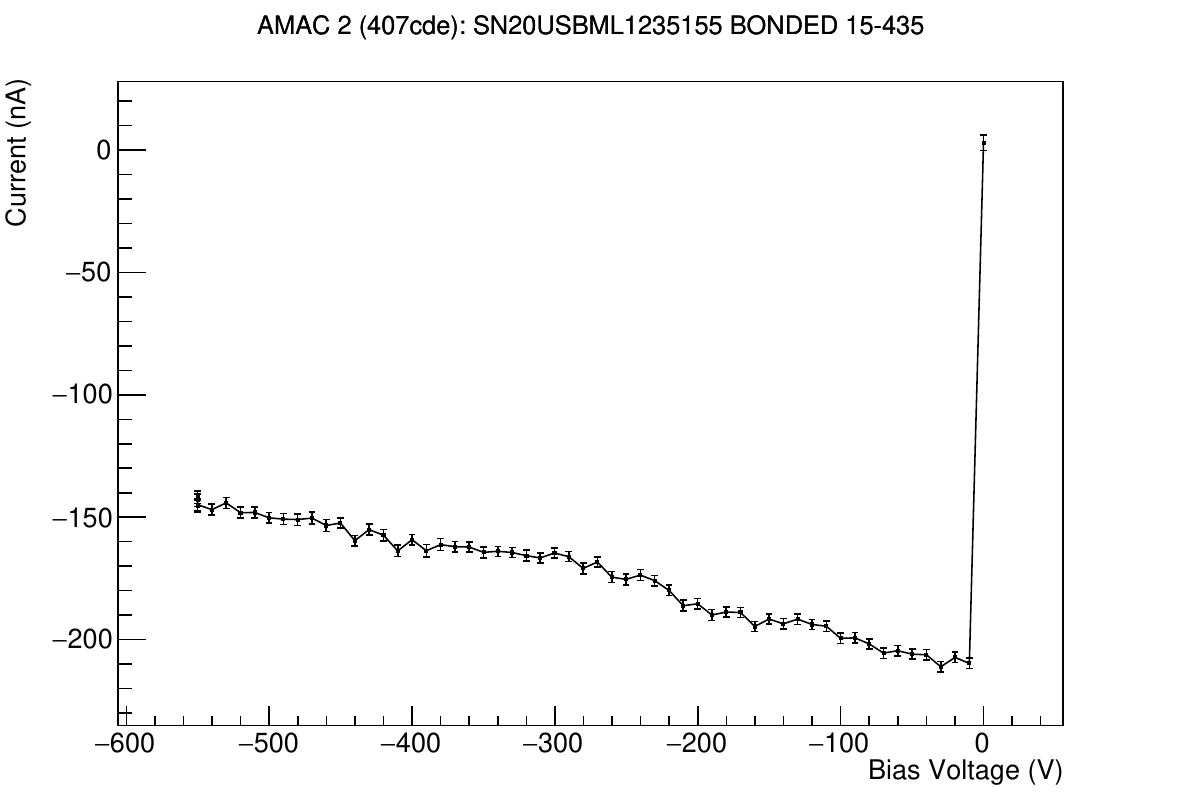}%
  }\hspace{-0.05\textwidth}
  \subfloat[Normal Module]{%
    \includegraphics[width=0.525\textwidth]{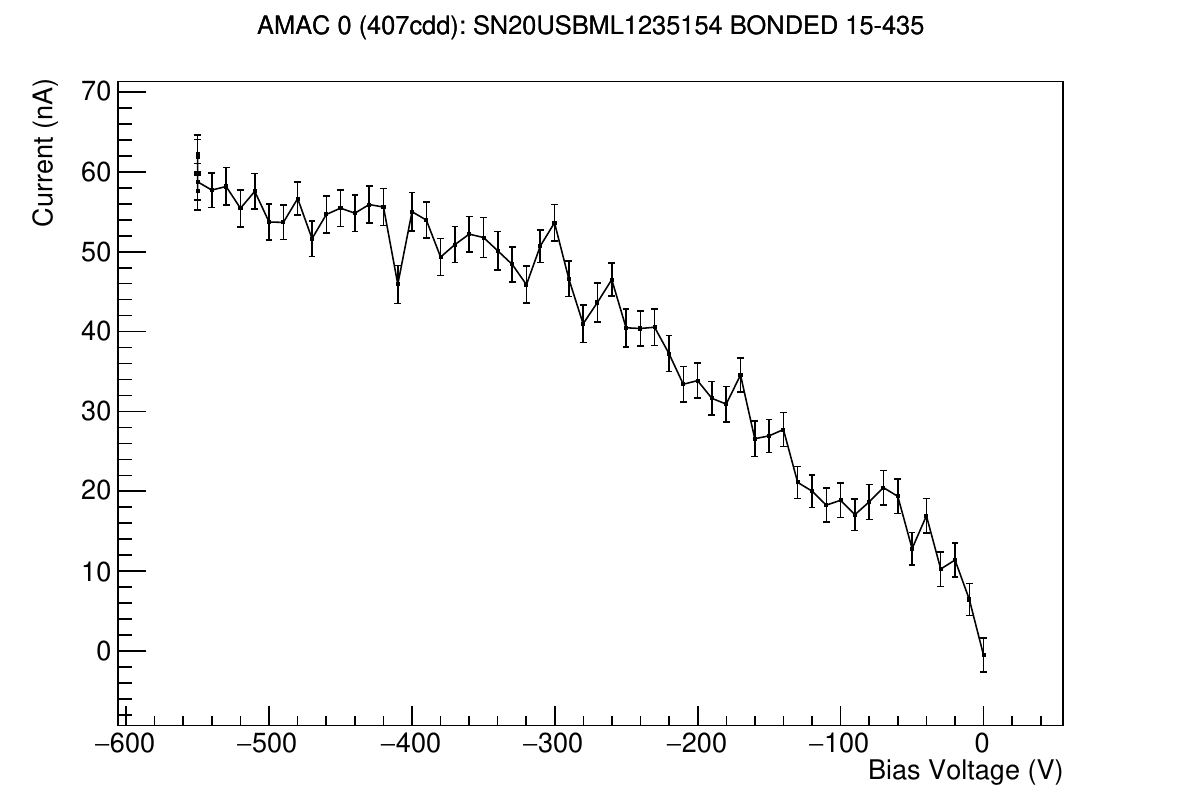}%
  }
\caption{The IV curve taken with unpowered ABCs, for (a) a module with pinholes observed at SCIPP, and (b) a more typical module with no pinholes. In the module with pinholes, current decreases between the 0V and -10V bias measurements, resulting in negative current values when the 0V bias current is subtracted from subsequent measurements.}
  \label{fig:negative_IV}
\end{figure}

At SCIPP, module SCIPP-LS-0091 was observed to show a negative current in the IV scan. With unpowered ABCs, a large negative jump was observed in the current between 0 and 10V, followed by the expected gradual increase of the leakage current as a function of bias voltage (Figure \ref{fig:negative_IV}). This indicates that the current at 0V, which is subtracted from subsequent current measurements, is higher than current measurements when bias voltage is applied.

Decreased AMAC current measurements with increased bias voltages can occur due to a change in direction of current from the ABC when small reverse bias voltages are applied. With no bias voltage applied, the ABC sinks current due to the relatively higher voltage of the AMAC non-inverting input. When a small bias voltage is applied, this current can decrease or even change directions, reducing the amount of current on the HV-return line.

Notably, modules with negative current values in unpowered ABC IV tests can yield normal IV results by starting the scan from -1V bias instead of 0V. This is because the algorithm subtracts the first current measurement in the voltage sweep from subsequent measurements, and measuring at an initial nonzero voltage reduces the change in this initial current due to interactions with the ABC.




\clearpage
\section{Locating pinhole-like connections}
\label{sec:LocatingPinholes}

When leakage current is not sufficiently high to affect the gain measured by the ABC, it can be difficult to locate a pinhole through the electrical channel calibration, especially if there are no visible defects in the sensor. Two methods were employed to magnify the effect of pinholes, with the aim of identifying the location of pinholes through subsequent module measurements.

\subsection{Modifying ABC input voltage}

By varying the internal bias voltage meant to tune the ABC preamplifier input transistor calibration, it is possible to change the ABC input voltage and increase or reduce the effect of pinholes. A test was developed in which all chips were set to the maximum internal bias setting, giving the highest possible ABC input voltage. Then, chip by chip the setting was reduced to its minimum, giving the lowest possible voltage. An increase in AMAC counts after adjusting this setting for a given chip indicates that the chip is connected to strips with pinholes.

\begin{figure}
\centering
\hspace{-12pt}\includegraphics[width=15 cm]{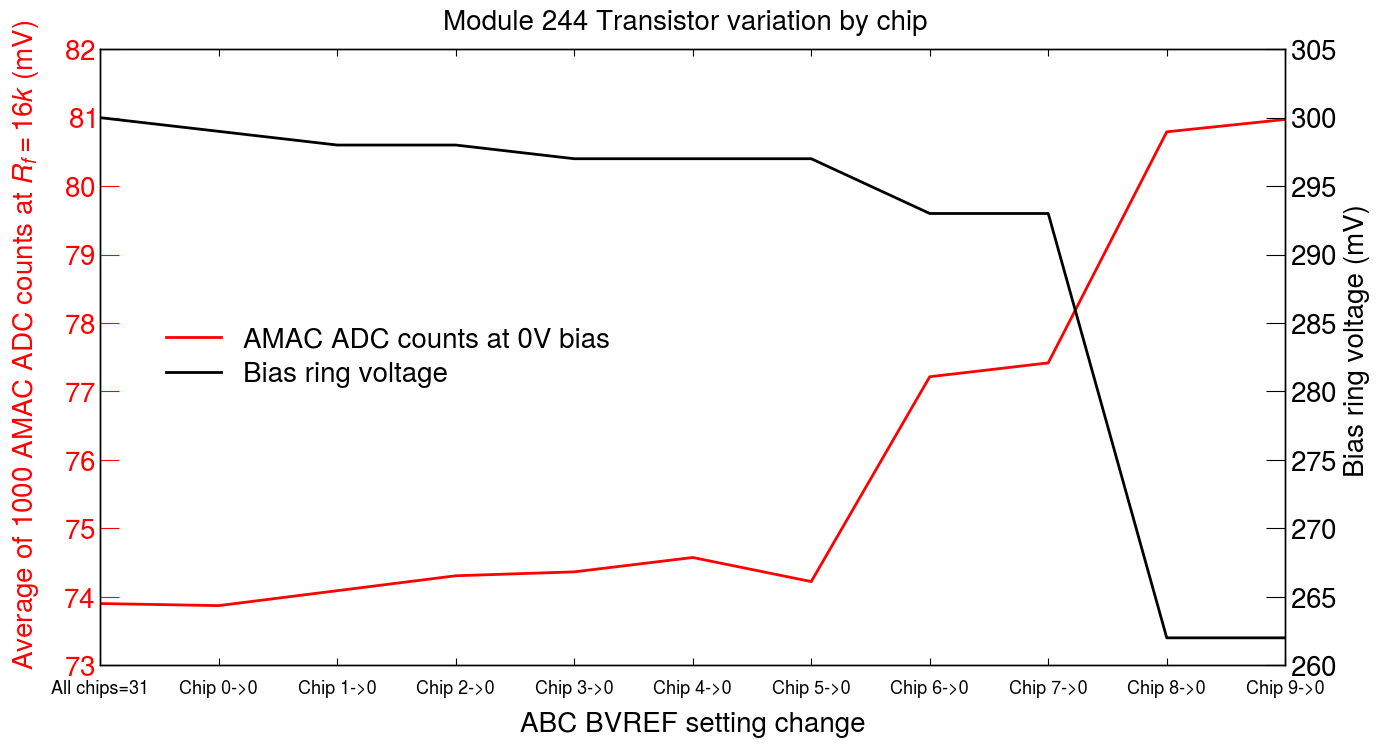}

\caption{Average of 1000 AMAC ADC counts (left axis) and bias ring voltage (right axis) on module BNL-PPB-MLS-244 as each ABC chip's internal bias voltage, BVREF, is decreased from the maximum setting to minimum setting. This decreases the ABC's input voltage, decreasing the voltage of the bias ring and increasing AMAC counts when the ABC is connected to a strip with a pinhole.}
\label{fig:ABCref}
\end{figure}

This test was first done at BNL on module BNL-PPB-MLS-244, which was modified to include a bond from the bias ring to an isolated pad on the module test frame, allowing simultaneous measurements of the bias ring voltage and AMAC counts. This module exhibited especially strong pinhole connections, with lowered AMAC counts at 0V bias even at a $16$ k$\Omega$ feedback resistance. Therefore, the test was performed at this lowered feedback resistance so that the AMAC HV-return measuring op amp would not saturate and a change in counts could be observed. At each chip setting, the bias ring voltage and the average of 1000 AMAC ADC counts were recorded (see Figure \ref{fig:ABCref}). The bias ring had a maximum voltage of 300mV with all chips set to the maximum internal bias voltage and 260mV with all chips set to the minimum internal bias voltage. When ABCs 6 and 8 are set to the minimum internal bias setting, the bias ring voltage sharply decreases and the AMAC counts sharply increase, indicating strong pinhole connections on these chips. Clear bonding issues were observed on both of these chips.

The same test was done at SCIPP for the SCIPP-LS-0091 module and is summarized in Table~\ref{tab:SCIPP-pinhole-counts} with $200$ k$\Omega$ feedback resistance. The large jump in counts when chips 0 - 2 are set to minimum voltage and when chips 0 - 3 are set to minimum voltage indicates that the pinhole is located on chip 3.

\begin{table}[h]
    \centering
    \begin{tabular}{lcccccccccc}
        \toprule
        All max & 0 & 1 & 2 & 3 & 4 & 5 & 6 & 7 & 8 & All min \\  
        \midrule
        56.79 & 56.68 & 56.49 & 56.52 & 62.22 & 62.18 & 62.11 & 62.00 & 62.02 & 62.02 & 61.92 \\  
        \bottomrule
    \end{tabular}
    \caption{The average of 1000 AMAC counts on SCIPP-LS-0091 when the internal bias voltage of all ABCs is set to maximum, and as each ABC is successively set to minimum internal bias voltage. Numbers 0 - 8 refer to the numbering of ABC chips on the module. Feedback resistance is set to $200$ k$\Omega$. The large jump between when chips 0 - 2 are set to minimum voltage and when chips 0 - 3 are set to minimum voltage indicates the pinhole is located on chip 3.}
    \label{tab:SCIPP-pinhole-counts}
\end{table}

While individual pinholed channels can be identified through front-end measurements when leakage current per strip exceeds 250 nA \cite{Affolder_2021}, this current does not occur during normal module and stave QC testing. Thus, this test provides an easy way to identify at least the ABC chip connected to a pinholed channel, using the nominal module and stave testing configurations. 

\subsection{Measuring gain during light exposure}

Leakage current is not sufficiently high to affect front-end measurements during normal testing conditions on a typical assembled module. However, the leakage current naturally increases due to irradiation during operations. It is possible to reach currents significant enough to impact the front-end operations by either exposure to the higher than nominal fluence or warmer temperatures at nominal end of life fluence \cite{Affolder_2021}. Alternatively, 
one can apply light exposure until gain measurements for channels with pinholes are noticeably decreased. Unlike the previous method, this technique potentially allows the identification of individual channels with pinholes. To test this, the leakage current was artificially increased by shining a light source on the sensors. The light source was oriented so that the coverage of the strips would be more or less uniform. The light intensity was adjusted to give values of 300 uA - 1000 uA for the total sensor leakage current. With 2560 strip channels and assuming uniform light illumination, this would therefore correspond to 117 - 390 nA per strip on average. The strip currents were further corrected for the degree of light screening of a strip by the flex circuits on top, which moved the maximum value in this range to over 500 nA for fully exposed strips.

\begin{figure}
\centering
\subfloat[]{\includegraphics[width=0.6\textwidth]{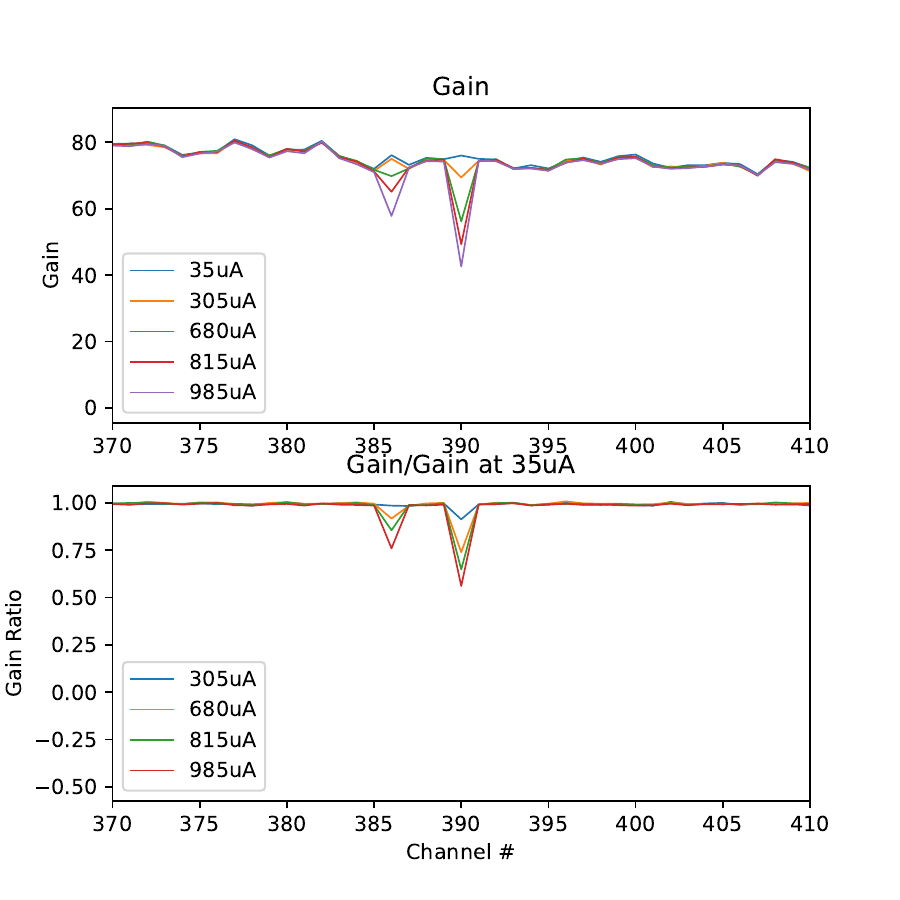}}\\
\subfloat[]{\includegraphics[width=0.8\textwidth]{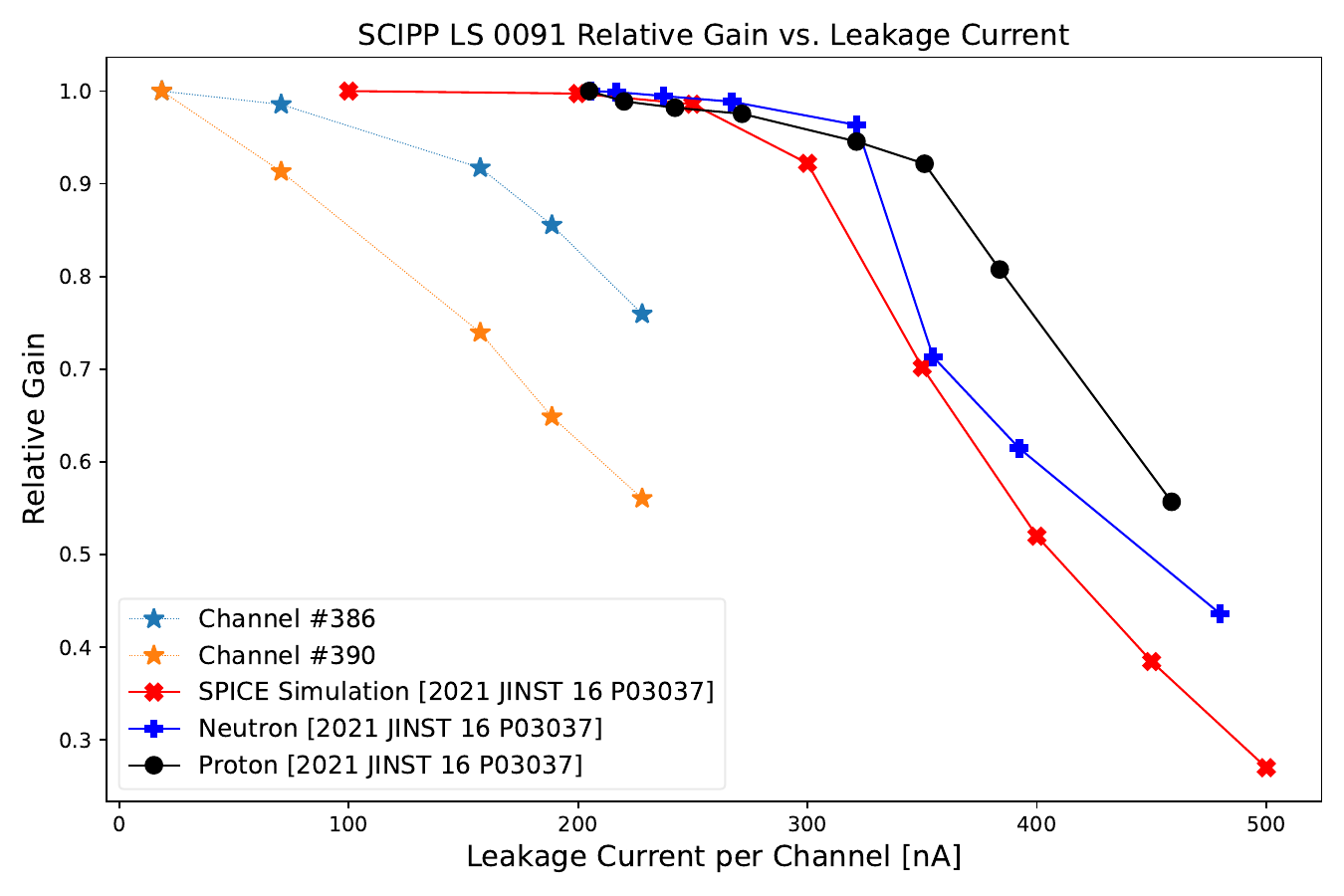}}

\caption{The strip gain as a function of the induced leakage current for SCIPP-LS-0091: (a) absolute and relative gain for different total sensor currents for consecutive channel numbers,  (b) gain versus per-channel current for the two affected channels. Plot (b) has been overlaid with gain values from a simulated pinhole setup modeled in SPICE and neutron and proton irradiated sensors (Affolder et al. \cite{Affolder_2021}). The measurement data from Affolder et al. has been adjusted to account for the different level of HV return potential due to AMAC reference voltage in the modules with powerboards.}
\label{fig:gain_vs_current_91}
\end{figure}

Out of the 2560 channels, only two channels, 386 and 390 exhibited a dependence on the leakage current, with a characteristic gain reduction at higher current levels (Figure \ref{fig:gain_vs_current_91}). Both channels were located on chip 3, in agreement with the previous tests from switching the ABC input calibration. 

\begin{figure}
\centering
\includegraphics[width=0.5\textwidth]{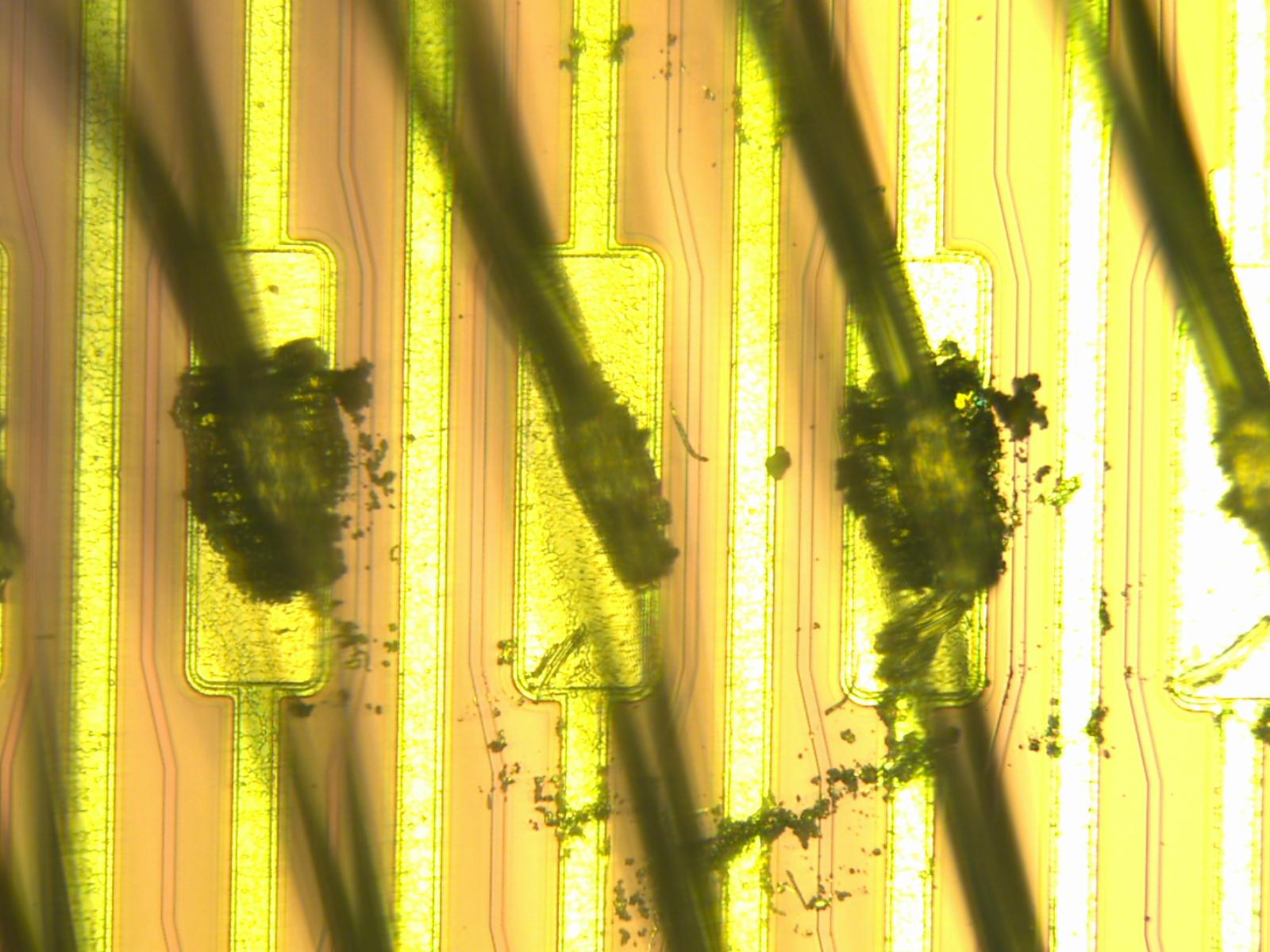}

\caption{Microscope image of two suspected pinhole channels on module SCIPP-LS-009, showing potential damage from contact from the bonding wedge with the silicon passivation layer. The wedge footprint appears to extend past the bonding pad and into the passivation later.}
\label{fig:tool_marks}
\end{figure}   

\begin{figure}
\centering
\includegraphics[width=0.7\textwidth]{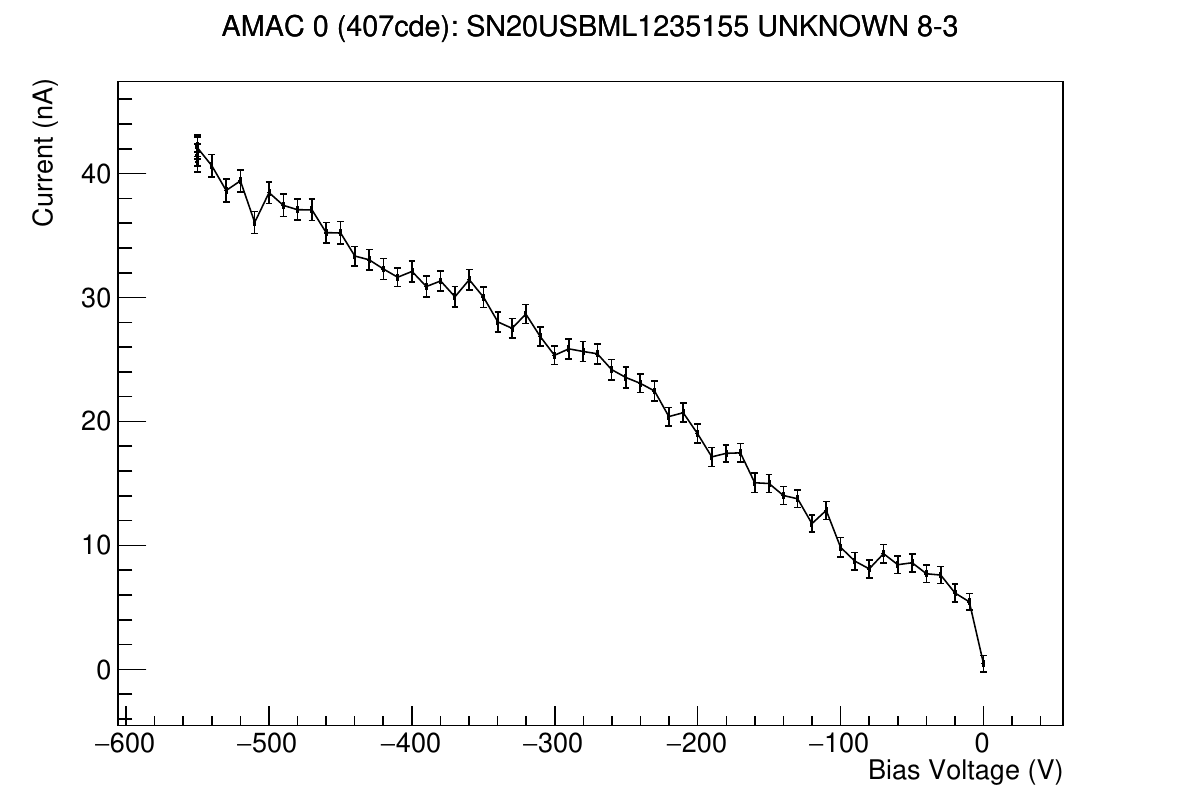}

\caption{The IV curve taken with unpowered ABCs for the module with pinholes at SCIPP after channels 386 and 390 were pulled, to be compared with the IV curve before the channels were pulled (Figure~\ref{fig:negative_IV}(a)).}
\label{fig:IV_before_after_91}
\end{figure}

Visual assembly showed tool marks around the two suspect channels which may have caused the pinhole (Figure \ref{fig:tool_marks}). During module assembly, it was recorded that the wire had come out of the bonding wedge prior to contact with the sensor and the channels had been subsequently rebonded. The wedge footprint is significantly larger than the nominal wire foot, as seen in the image. Without the wire in the wedge, the impact of the tool could have broken the passivation layer and created a parasitic connection between the metal and implant strips, resembling a pinhole. After the two bonds were pulled, all symptoms of pinholes disappeared. The AMAC IV looked normal again and there were no jumps in the AMAC counts (Figure \ref{fig:IV_before_after_91}). 

\begin{figure}
\centering
\includegraphics[width=0.8\textwidth]{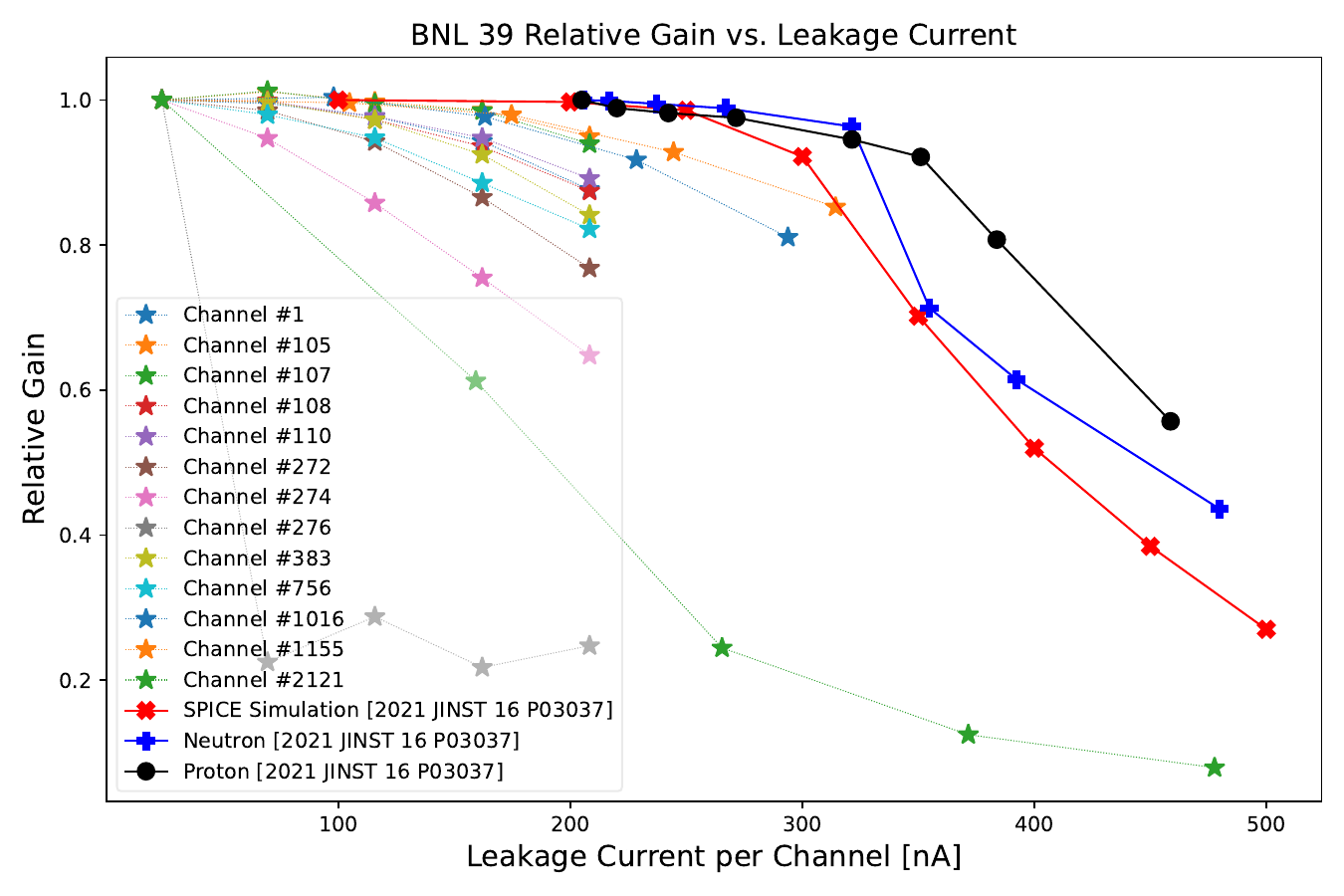}
\caption{Relative gain as a function of strip leakage current for module BNL-LS-39. The plot has been overlaid with gain values from a simulated pinhole setup modeled in SPICE and neutron and proton irradiated sensors (Affolder et al. \cite{Affolder_2021}). The measurement data from Affolder et al. has been adjusted to account for the different level of HV return potential due to AMAC reference voltage in the modules with powerboards.}
\label{fig:bnl_39}
\end{figure}

A similar set of tests was done for another module, BNL-LS-39. In this case, many more pinholes were identified. The pinhole detection method of scanning the ABC preamplifier settings did not yield the locations, due to their spatially wide distribution. However, the light test detected them via 
the same trend for gain vs leakage current (Figure \ref{fig:bnl_39}). Again, visual inspection showed potential damage at the pinhole channels with wide bonding wedge footprint. The corresponding front-end wire bonds were subsequently pulled and the pinhole features disappeared. The increased statistics of this case confirm that there is a significant spread of the gain dependence on current for different channels, as already indicated in Figure \ref{fig:gain_vs_current_91}.

\section{Artificially inducing pinholes}
As a subsequent test, SCIPP tried to artificially create pinholes on two modules. The front-end bonds to ABC chips were kept in place, and the bonding pads on the opposite strip end were impacted with a bonding wedge with no wire. The bonding process was standard: the wedge was pressed down on the pad with specified force for the bonding time duration while ultrasonic energy was applied. Under the standard conditions, with the bondwire present, this process squashes the wire, making a "foot" on the pad surface, and welds it to the pad metal. In our tests without the wire the (wider) wedge interacts with the pad directly, making an imprint and possibly affecting the material under the metal.

The tested pad locations were recorded and subsequently imaged. The light tests, pinhole detection by ABC setting scan, and IV tests were performed. In the following, the pinhole creation attempt was deemed successful when gain affected by the strip current changed during the light test.

Under nominal bonding parameters the success rate in creating the pinholes was very high. We varied the bonding force in the range of 16-30 cN, including the standard value of 22 cN. Pinholes were created in all cases, although the physical damage was visually more significant for higher force values.

\begin{figure}
\centering
\subfloat[]{\includegraphics[width=0.36\textwidth]{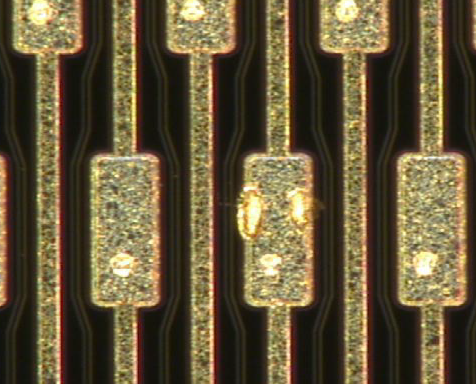}}
~~\subfloat[]{\includegraphics[width=0.3\textwidth]{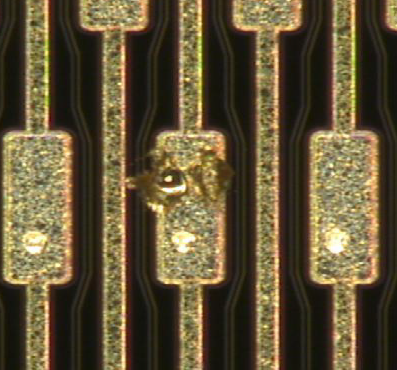}}
\caption{Visual inspection of the pads after the bonding wedge touch-down: (a) after 5 ms bonding time, (b) after 45 ms bonding time.}
\label{fig:vi_damage}
\end{figure}

\begin{figure}
\centering
\includegraphics[width=0.5\textwidth]{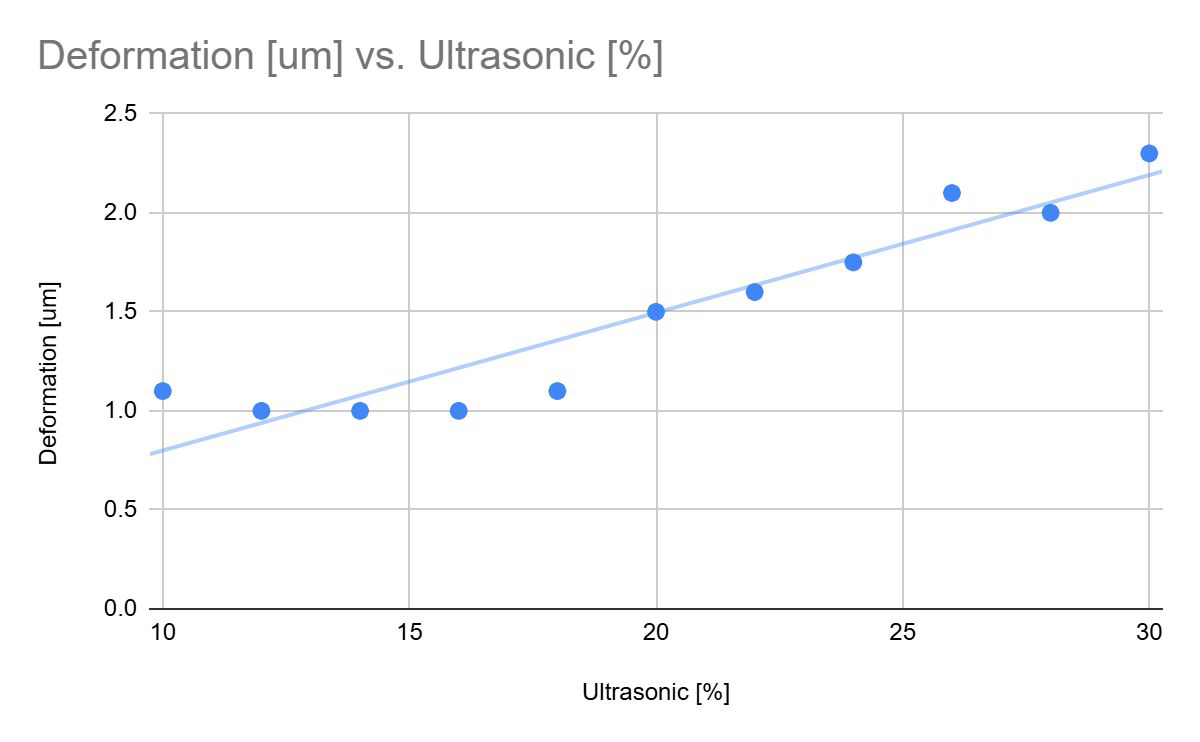}

\caption{Surface deformation reported by wirebonding as a function of ultrasonic energy.}
\label{fig:deform_us}
\end{figure}

Similarly, we performed custom tests and scans of several other parameters:
\begin{itemize}
    \item Bond time was varied in the range of 5-45 ms, compared to the standard time of 50 ms. The visual damage was more obvious at higher values (Figure \ref{fig:vi_damage}), and there was a correlation of pad deformation\footnote{Pad deformation is defined as the displacement of the bonding wedge and pad perpendicular to the sensor. It measured as a function of time by the wirebonding machine.} with bond time. The light test indicated pinholes for times $\geq$ 40 ms, when deformation value was $\geq$ 1.2 $\mu$m.
    \item Ultrasonic energy was scanned and also found to correlate with the degree of the estimated surface deformation, exhibiting a threshold behavior for the pinhole creation success (Figure \ref{fig:deform_us}). Note that thickness of the aluminum wirebonding pad is $< 1\mu$m.
    \item Wedge angle  with respect to the pad was varied between 0 and 10 degrees. All attempts were equally successful in creating the pinhole connections.
    \item In some cases of unsuccessful pinhole creation attempts, bond feet were added to imitate the re-bonding process on real modules in case of wire loss. This addition did not result in a pinhole connection. This observation confirms that the pinholes were created by the ultrasonic energy applied to the wireless wedge after the initial touch-down, not during re-bonding.
    \item Scratches on the sensor surface were added with metal tweezer tips. This did create pinholes, however scratch symptoms were clearly visible in other readout tests, indicating that the channels involved had issues beyond pinholes.
\end{itemize}

\begin{figure}
\centering
\subfloat[]{\includegraphics[width=0.8\textwidth]{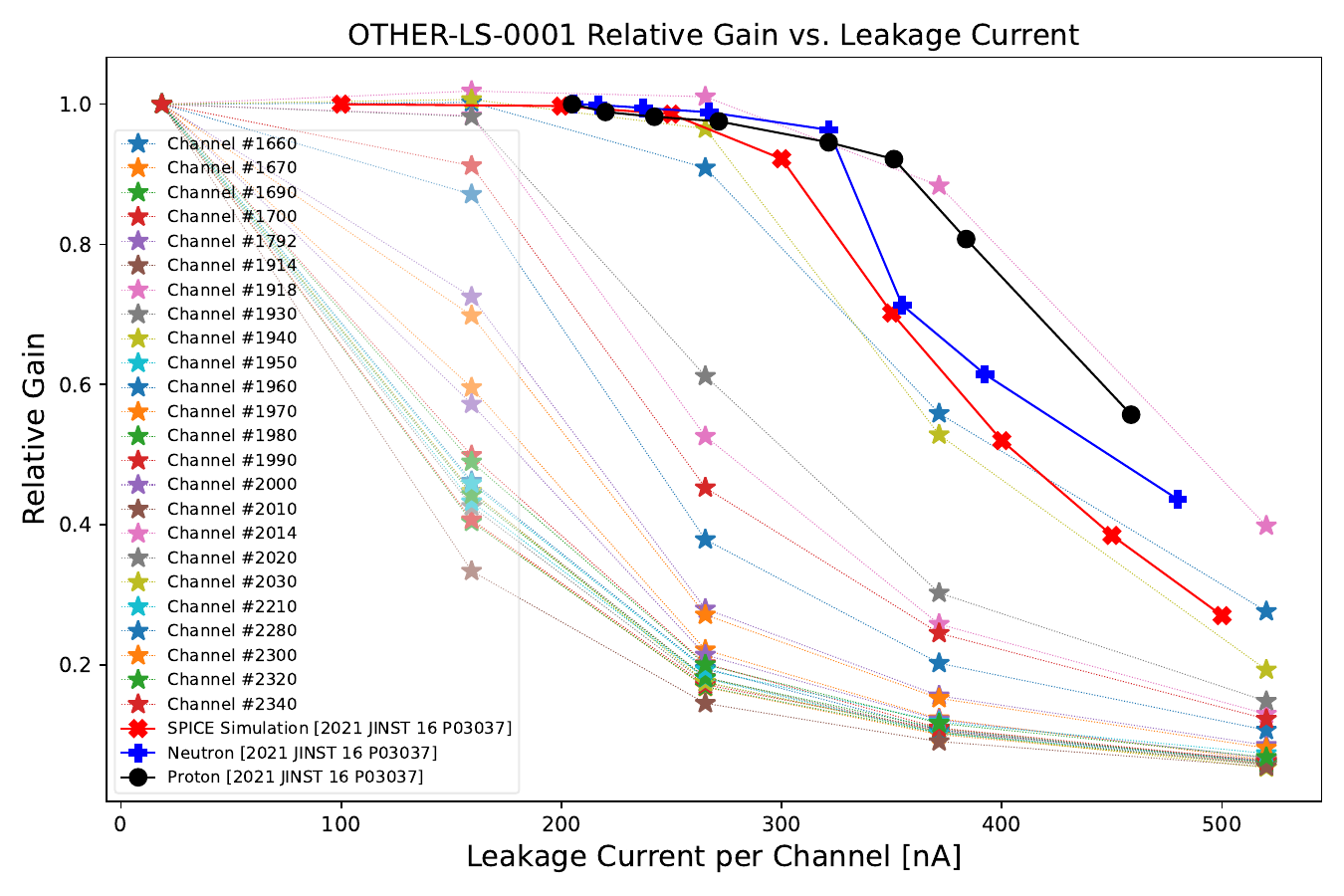}}\\
~~\subfloat[]{\includegraphics[width=0.8\textwidth]{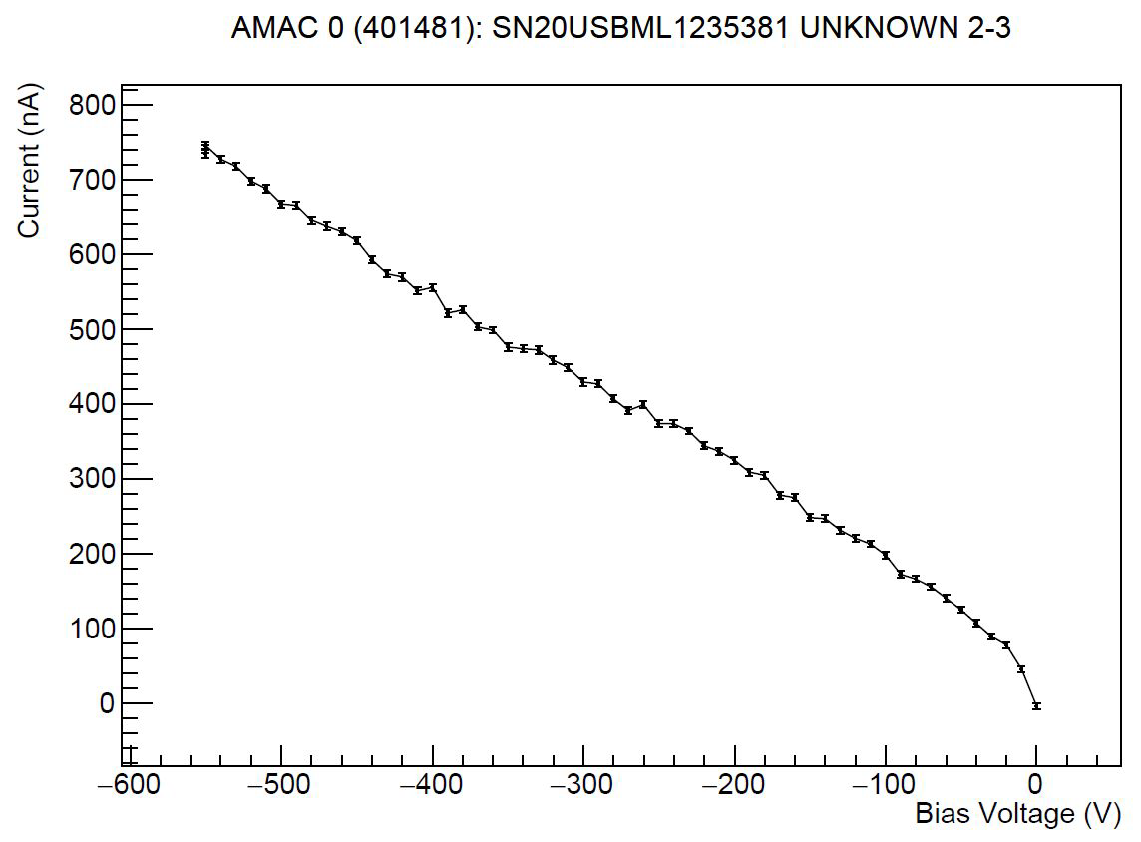}}
\caption{Test results for a module with intentional pinhole creation by bonding: (a) Gain as a function of channels with pinhole connections. The plot has been overlaid with gain values from a simulated pinhole setup modeled in SPICE and neutron and proton irradiated sensors (Affolder et al. \cite{Affolder_2021}). The measurement data from Affolder et al. has been adjusted to account for the different level of HV return potential due to AMAC reference voltage in the modules with powerboards. (b) IV plot with unpowered ABCs. }
\label{fig:art_damage}
\end{figure}

The gain characteristics for the affected channels resembled the behavior observed for module BNL-LS-39, including the significant channel-to-channel variability (Figure \ref{fig:art_damage}). This confirms that wire loss during the wire bonding process is a valid pinhole creation mechanism. Note that the IV test shows normal behavior with no negative current values, indicating that negative values are not a reliable indicator of pinholes. 

Additionally, a fraction of channels in Figure \ref{fig:art_damage} exhibit decreased gain at lower currents than anticipated. The inconsistency in onset current with the previous study \cite{Affolder_2021}, which simulated pinholes with wirebonded shorts on undamaged strips, is likely due to variability in the properties of the cracked coupling capacitor or a local strip current increase due to the damage, compared to the estimated value. This new data appear to contain two distinct behaviors. Some channels are fairly consistent with the results from the prior study, while others have a much earlier gain reduction, indicative of the additional damage.

\section{Modifications to testing and assembly procedures}

Due to the discovery of pinholes and their effect on leakage current measurements in fully completed modules, standard IV scan procedure was changed to turn off the ABCs before the scan. This ensures AMAC counts are not lowered or saturated because of pinholes. IV scans taken with unpowered ABCs may show negative currents due to 0V offsets that are higher than measurements at higher bias voltages. To prevent negative currents, one can take an IV scan where the initial 0V offset is actually measured at a small bias voltage such as -1V so that all measurements are taken at a constant ABC input voltage. This can be done to verify the accuracy of IV scan results or confirm the negative current was the result of a pinhole. Regardless of whether the 0V offset is measured at 0V or a small bias voltage, early breakdown behavior is clearly identifiable when IV scans are taken with the ABCs turned off.

Figures \ref{fig:gain_vs_current_91}, \ref{fig:bnl_39}, and \ref{fig:art_damage} show significant variation in the onset of the gain reduction, with a fraction of channels exhibiting decreased gain at lower currents than seen by Affolder et \cite{Affolder_2021}. These currents approach the expected end-of-life leakage currents in the detector. As a result of the discrepancy in gain reduction onsets, the wirebonding procedure now emphasizes that strips damaged during bonding should not be re-bonded in order to prevent unacceptable levels of unexpected bad strips at the end of the detector's lifetime.

The methods described in Section \ref{sec:LocatingPinholes} to locate pinhole-like connections are not included in the standard module testing procedure. However, module testing sites are encouraged to employ these methods whenever results from regular QC tests are not fully understood. For example, after obtaining a negative IV test, a module site may wish to measure gain as a function of leakage current with some light exposure. This could reveal a channel with bonding damage that was not noted during the wirebonding procedure.

\section{Conclusion}

We have studied the effect of pinholes on HV-return measurements in ITk strip modules. Pinholes have been observed via such measurements during module and stave production across several institutions and in both barrel and endcap modules. It was determined that pinholes introduce a DC connection between front-end chips and the HV-return, which increases or decreases currents measured by front-end electronics depending on whether the front-end is powered. Thus, thorough understanding of pinhole effects of HV-return measurements is necessary to interpret IV scan results.

Sudden changes in the offset during an IV scan are a frequent, but not guaranteed, indicator of the presence of pinholes. In some cases, it is possible to locate the ABC chip connected to strips with pinholes by varying the ABC input voltage and measuring subsequent changes in AMAC ADC counts. This method is especially effective when sensor damage is localized but may fail if the pinholes are widely distributed among multiple chips.  The most precise determination of pinhole locations is by increasing sensor current either by exposure to light or, ultimately, irradiation and looking for a decrease in ABC channel gain. These tests are proposed as additional diagnostics for modules and staves whose results are not fully understood. The identification and location of pinholes can provide valuable insight into sensor damage that occurs during Strip detector production. For example, we have shown pinholes can point to wirebonding damage and indicate the formation of a crack in the sensor. 

Pinholes naturally occur very rarely in the sensors. The previous study \cite{Affolder_2021} dealt with several channels simulated with wirebonding shorts. In this work we diagnosed pinholes that can occur due to wirebonding process issues with significantly expanded statistics on several modules, including their intentional reproduction at different bonding conditions. Notably, the same qualitative gain deterioration with the strip current was seen. However, the onset of the gain reduction varies significantly and occurs at lower currents than anticipated. The gain behavior was not correlated with bonding force or other studied parameters. This inconsistency is attributed to differences between pinholes simulated by wirebonding shorts and pinholes created through damage and is addressed in the module wirebonding procedure.

IV scanning procedures were altered to minimize the effect of pinholes during QC tests. With the ABCs unpowered during scanning, pinholes do not interfere with the detection of early breakdown. Increased 0V offset due to pinholes can result in negative current measurements, but similarly does not interfere with early breakdown detection as the magnitude of current increase is preserved. The offset can be removed by starting the scan at -1V instead of 0V, which becomes an additional diagnostic method. We conclude that with these modifications, pinholes do not pose a risk to quality of module current measurements throughout detector operations. 

\acknowledgments
Work performed at SCIPP was supported by the US Department of Energy, grant DE-SC0010107.

Work performed by Emily Duden was supported by the U.S. Department of Energy, Office of Science, Office of Workforce Development for Teachers and Scientists, and Office of Science Graduate Student Research (SCGSR) program. The SCGSR program is administered by the Oak Ridge Institute
for Science and Education (ORISE) for the DOE. ORISE is managed by ORAU under contract number DE-SC0014664. All opinions expressed in this paper are the author’s and do not necessarily reflect the policies and views of DOE, ORAU, or ORISE.

The authors gratefully acknowledge help from Matthew Gignac for providing test data from the previous study.

\printbibliography





\end{document}